\documentclass{emulateapj}
\usepackage{epsfig}
\usepackage{amsmath,amssymb}

\newcommand{\hMsol}{{\>h^{-1}\rm M}_\odot}
\newcommand{\hGpc}{{\>h^{-1}\rm  Gpc}} 
 
\newcommand{\hMpc}{{\>h^{-1}\rm  Mpc}} 
\newcommand{\hkpc}{{\>h^{-1}\rm kpc}}

\begin{document}
\title{Anisotropy in the matter distribution beyond the baryonic\\acoustic oscillation scale}
\author{
  A. Faltenbacher\altaffilmark{1} and 
  Cheng Li\altaffilmark{2} and 
  Jie Wang\altaffilmark{3}}
\altaffiltext{1}{Physics Department,University   of   the   Western   Cape, Cape
  Town   7535,   South Africa}
\altaffiltext{2}{Partner Group of the MPI f\"{u}r Astrophysik at
  Shanghai Astronomical Observatory and Key Laboratory  for Research 
  in  Galaxies and Cosmology  of Chinese  Academy  of  Sciences,
  Nandan   Road  80,  Shanghai  200030,  China} 
\altaffiltext{3}{Institute for Computational Cosmology, Department of Physics,
  University of Durham, South Road, Durham DH1 3LE }
\begin{abstract}
  Tracing the  cosmic evolution  of the Baryonic  Acoustic Oscillation
  (BAO) scale with galaxy two point correlation functions is currently
  the most promising approach to detect dark energy at early times.  A
  number of ongoing  and future experiments will measure  the BAO peak
  with  unprecedented accuracy.   We show  based  on a  set of  N-Body
  simulations  that  the matter  distribution  is  anisotropic out  to
  $\sim150\hMpc$,  far beyond  the  BAO scale  of $\sim100\hMpc$,  and
  discuss implications for the measurement of the BAO. To that purpose
  we  use  alignment correlation  functions,  i.e., cross  correlation
  functions  between  high  density   peaks  and  the  overall  matter
  distribution  measured  along  the  orientation  of  the  peaks  and
  perpendicular  to  it.   The  correlation  function  measured  along
  (perpendicular to) the orientation of high density peaks is enhanced
  (reduced)  by  a  factor   of  $~2$  compared  to  the  conventional
  correlation function and the location of the BAO peak shifts towards
  smaller  (larger) scales  if measured  along (perpendicular  to) the
  orientation of the high density peaks.  Similar effects are expected
  to shape observed galaxy correlation functions at BAO scales. 
\end{abstract}
\section{Introduction}
\label{sec:intro}
The  Baryonic Acoustic  Oscillations (BAO) constitute a characteristic
feature within the large scale structure of the Universe and can serve
as  standard ruler  for  constraining the  properties  of dark  energy
\citep[e.g.,][]{Blake-Glazebrook-03, Linder-03,   Seo-Eisenstein-03,
  Wang-06,   McDonald-Eisenstein-07, Seo-Eisenstein-07, Seo-08,
  Seo-09, Kazin-Sanchez-Blanton-12}.  The BAO in the  baryon-photon fluid of  the
pre-recombination era  imprint the sound horizon distance at
decoupling  as a typical scale in the matter correlation   function
or   power   spectrum   \citep{Peebles-Yu-70, Sunyaev-Zeldovich-70,
  Eisenstein-Hu-99, Bashinsky-Bertschinger-02}. These oscillations
were detected  in the cosmic  microwave background
\citep[e.g.,][]{Page-03} and  in the spatial  distribution of galaxies 
\citep{Eisenstein-05, Cole-05} and have  been confirmed by a number of
subsequent  studies  \citep[e.g.,][]{Percival-07,  Cabre-Gaztanaga-09,
  Sanchez-09, Percival-10, Reid-10, Kazin-10}.

The next generation of large galaxy surveys, like the Panoramic Survey
Telescope \& Rapid  Response System (Pan-STARRS, \citealt{Kaiser-02}),
the   Dark  Energy  Survey   (DES,  \citealt{DES-05}),   the  Baryonic
Oscillation Spectroscopic Survey (BOSS,
\citealt{Schlegel-White-Eisenstein-09}), BigBOSS \citep{Schlegel-11}, 
the   Hobby   Eberly  Telescope   Dark   Energy  Experiment   (HETDEX,
\citealt{Hill-04})    and    the    space   based    Euclid    mission
\citep{Cimatti-09},  will  cover  volumes  much  larger  than  current
datasets, allowing for much more accurate determinations of the BAO.

In the two-point correlation function  the BAO are visible as a unique
broad  and  quasi Gaussian  peak  \citep{Matsubara-04}.  However,  the
determination of the shape and location of the peak may be affected by
sample variance \citep{Cabre-Gaztanaga-09, Martinez-09, Kazin-10,
  Cabre-Gaztanaga-11}, non   
linear effects  and the bias of  the tracer galaxy  population
\citep{Smith-03,  Crocce-Scoccimarro-06a, Angulo-08, 
  Crocce-Scoccimarro-08,    Seo-08}.     These    difficulties    lead
\cite{Prada-11} to suggest  to use the zero-crossing of  the two point
correlation located  at $\sim 130\hMpc$  as standard ruler  instead of
the  peak  location.   Yet,  several observations do not show  the
theoretically  predicted  zero-crossing at all \citep{  Martinez-09,
  Kazin-10}.  At this stage it is unclear whether this discrepancy is
caused  by  systematic  effects   or  cosmic  variance  or  whether it
represents  a   challenge  for  the   concordance  $\Lambda$CDM  model
\citep{Labini-09, Sanchez-09, Kazin-10}.

One  basic assumption for  interpretation of  the BAO  measurements is
that  the matter  distribution  is isotropic  at  the relevant  scales
($\sim100\hMpc$). In this work  we use alignment correlation functions
\citep{Paz-Stasyszyn-Padilla-08,   Faltenbacher-09}   to   show   that
the amplitudes of the two-point correlation function measured along the
orientations of the  high density peaks are larger  than those derived
from  spherically averaged (conventional)  clustering analysis  out to
scales of $\gtrsim 150\hMpc$ and discuss possible effects on
measurements of the BAO. 
\begin{figure*}[t]
  \epsfig{file=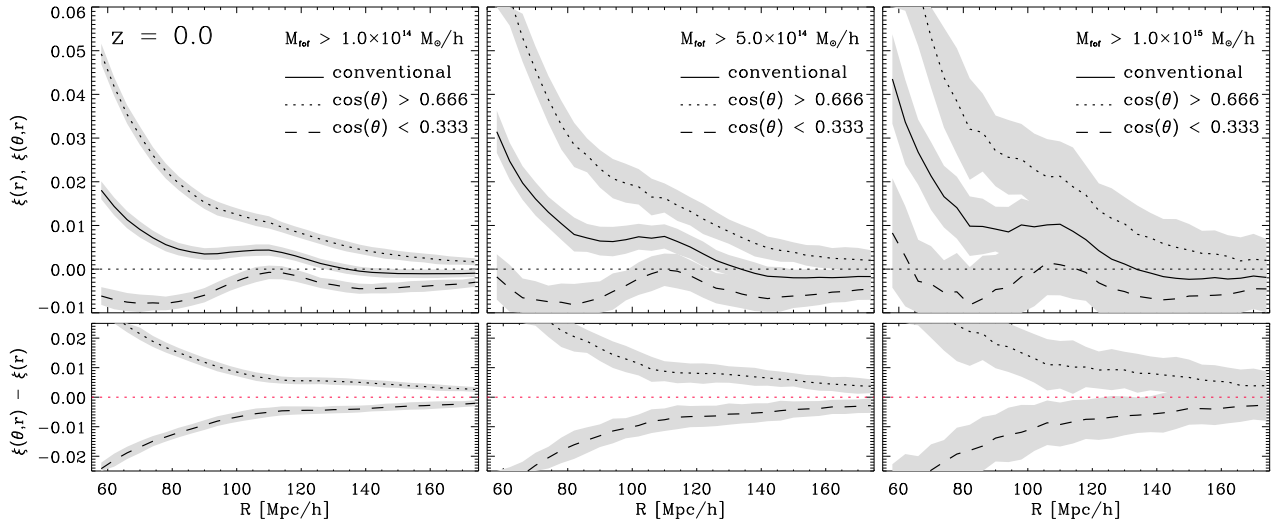,width=0.95\hsize }
  \caption{\label{fig:z0} {\it Upper  panels:} 3 dimensional two point
    cross  correlation functions  between FoF groups of  the indicated
    mass  and the  overall matter  density field  represented  by 10\%
    random subset  of all simulation particles.  Solid  lines show the
    {\em conventional} correlation functions.  Dotted and dashed lines
    show   the  {\em  alignment}   correlation  functions   along  and
    perpendicular to the orientations  of the FoF groups,
    respectively.   {\it   Lower   panels:}  Difference   between   
    conventional and alignment  correlation functions above.
    The shaded  areas in the upper and lower panels   give  the
    cosmic  variance.  Poisson  errors  are 
    negligible and are not shown here.}
\end{figure*}
\section{Methodology}
We use a  set of large scale dark matter  simulations which follow the
evolution  of the cosmic  density field  including its  collapse into
high density peaks.  Based on their orientations we
compute the alignment cross correlation function with the over all mass
distribution out to $\sim200\hMpc$. In this section we briefly discuss
the simulations, the computation of density peak orientations and the
definition of the alignment correlation function.
\subsection{Simulations} 
We used GADGET-2 \citep{Springel-05b} to carry out 50
$1\hGpc$-simulations based on the  same concordant  $\Lambda$CDM  cosmology  but  different
realizations of  the initial  density field. The  total volume  of the
ensemble is almost  70 times larger than the  volume used for the
detection of the BAO signal by \citet{Eisenstein-05}.   The
cosmological  parameters  correspond  to 
those used  for the {\it  Millennium Simulation} \citep{Springel-05a}:
matter density parameter, $\Omega_{\rm M}=0.25$; cosmological constant
term,    $\Omega_{\Lambda}=0.75$;   power    spectrum   normalization,
$\sigma_8=0.9$;  spectral   slope,  $n_s=1$;  and   Hubble  parameter,
$h=0.73$. For each run $320^3$  particles are used resulting in a mass
resolution  of  $2.11  \times  10^{12}  \hMsol$ for  the  dark  matter
particles. A Plummer-equivalent softening of $80\hkpc$ was employed.
An analysis based on a much lower force resolution showed very similar
results.
\subsection{Orientations of high density peaks} 
Density   peaks   are  identified   by   using  a   friends-of-friends
\citep[FoF,][]{Davis-85} group-finder with linking parameter $b = 0.2$
times  the mean  particle separation.  In each  realization  there are
about 30,000 FoF-groups with masses greater than $10^{14}\hMsol$.
The  orientations are  derived by  evaluating the  eigenvalues  of the
second  moment of the mass  tensor.  In the  following  the direction  of
principal axis  associated with the largest eigenvalue  is referred to
as  the (3D)  orientation  of the  density  peak.  Observationally  3D
orientations  are  difficult  to  determine, therefore,  we  also  use
projected orientations.   These are determined by  collapsing the mass
distribution  of  the density  peak  along  the  line of  sight  (here
equivalent   to  the   z-axis)  and,   consecutively,   computing  the
eigenvectors  of   inertia  tensor   for  the  two   dimensional  mass
distribution. Despite  the poor  force resolution the  orientations of
the density  peaks are expected to  trace the large  scale tidal field
well (see below).
\begin{figure*}
  \begin{center}
  \epsfig{file=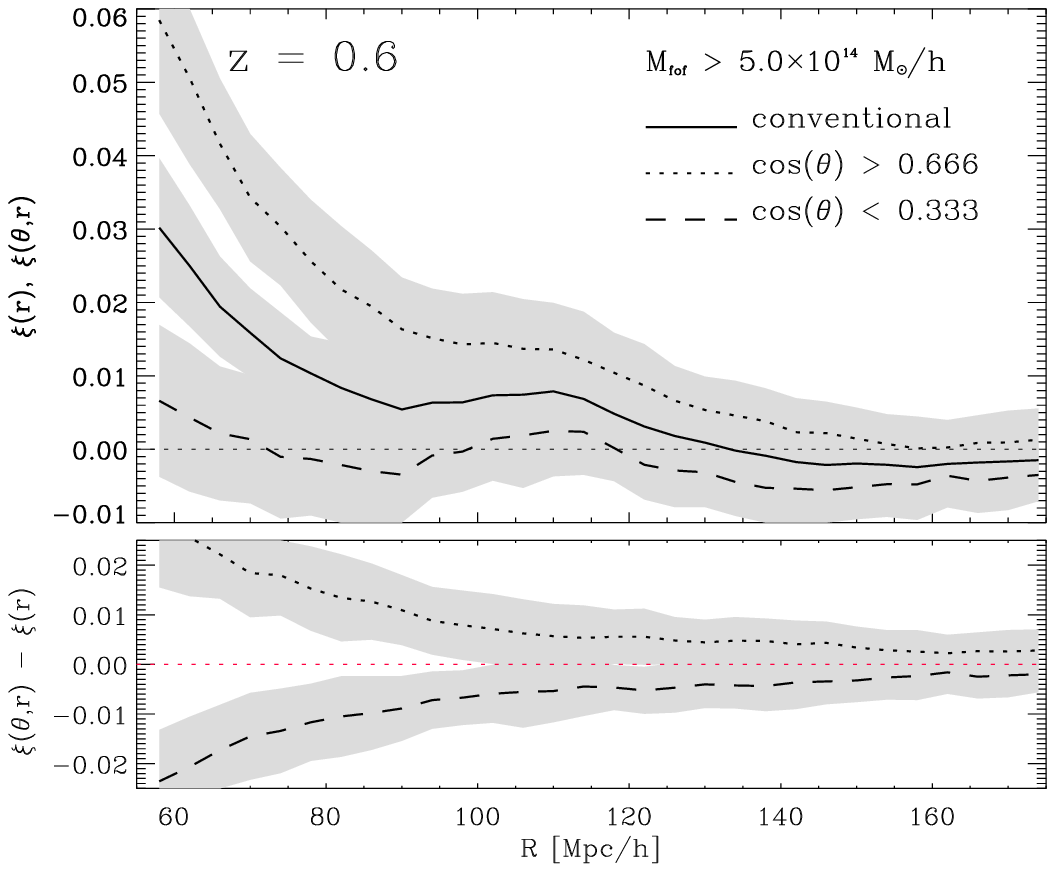,width=0.4\hsize }
  \hspace{0.09\hsize}
  \epsfig{file=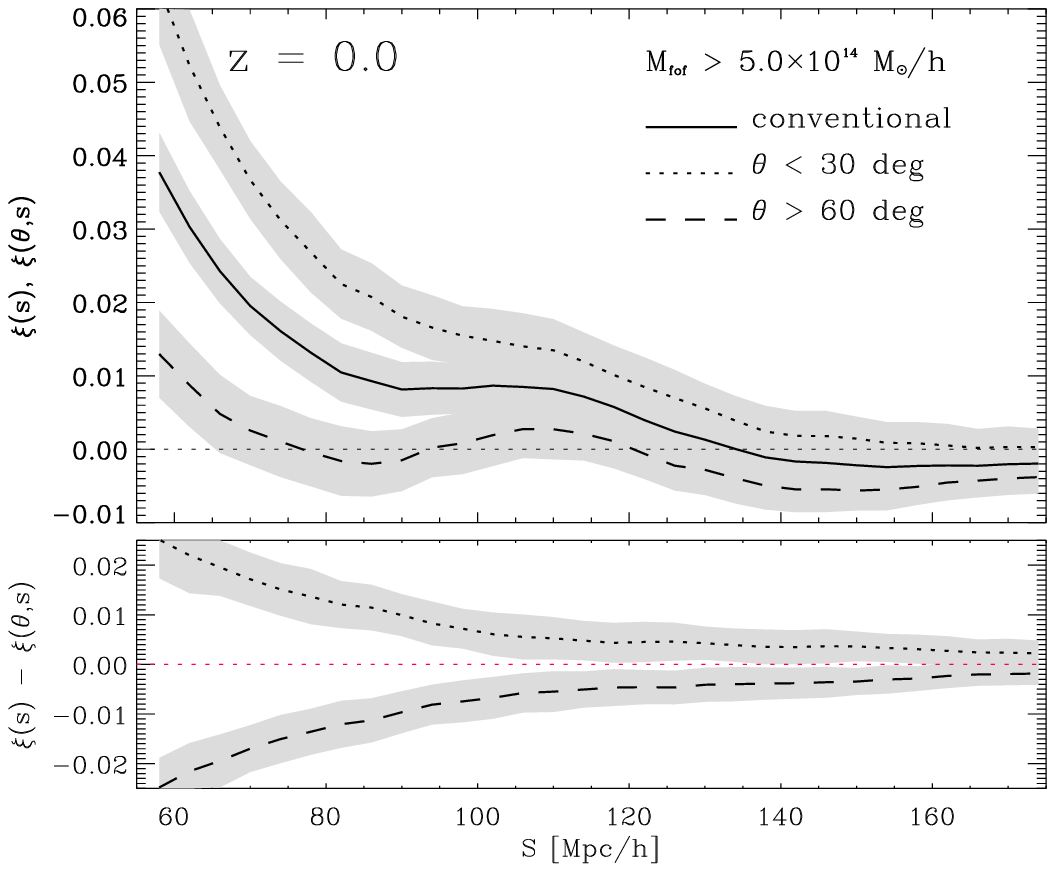,width=0.4\hsize }
  \end{center}
  \caption{\label{fig:z06}{\it Left panel:} Same  as middle panel in
    Fig.~\ref{fig:z0} for a redshift of $z=0.6$. {\it Right panel:}
    Same as middle panel in Fig.~\ref{fig:z0} but using projected
    orientations and redshift space separations (at $z=0$).}
\end{figure*}
\subsection{Alignment correlation functions} 
Two-point  correlation functions,  $\xi(r)$,  are a  primary tool  for
quantifying  the  clustering  properties  of  cosmological  structures
\cite[e.g.][]{Peebles-80}.We employ an extension of the {\em two-point
  cross-correlation function}  which can be used  to quantify spatial
alignment  of  objects in  one  sample  with  respect to  orientations
assigned     to     the     objects     in    the     other     sample
\citep[see,][]{Paz-Stasyszyn-Padilla-08,               Faltenbacher-09,
  Schneider-11}.  Given a sample  of objects in question (Sample $Q$),
a sample of reference objects (Sample $R$) and a random sample (Sample
$\mathcal{R}$)  we  can define  the  {\em  alignment  two point  cross
  correlation function}
\begin{equation}\label{eqn:xipv}
  \xi(\theta,r)=
  \frac{N_{\mathcal{R}}}{N_{R}}\frac{QR(\theta,r)}{Q\mathcal{R} 
    (\theta,r)} - 1,
\end{equation}
where $r$  indicates the distance,  $\theta$, is the angle  between the
orientation of  the objects in Sample  $Q$ and the  line connecting to
objects   in  either   of   the  two   other   samples.  $N_{R}$   and
$N_{\mathcal{R}}$ are the number  of points contained in the reference
and random sample, $QR(\theta,r)$ and $Q\mathcal{R}(\theta,r)$ are the
counts of cross pairs between the indicated samples for given $\theta$
and $r$.  The conventional correlation  function is the average of the
alignment correlation function over the full range of $\theta$ values.

Taking symmetries into account the value of $\theta$ ranges from zero
(along   the  orientation   of  the   main  object)   to   90  degrees
(perpendicular to  its orientation).   Thus, higher amplitudes  of the
alignment correlation  functions at small  values of $\theta$  indicate that
the reference objects are more likely to be aligned along orientation
of the main objects.  In contrast, higher amplitudes at larger angles
indicate that  the reference  objects are more  likely to  be located
perpendicular to the orientation of the main objects.

We  employ  the alignment  correlation  functions,  more exactly,  the
average  of the  alignment correlation  function within  given angular
ranges,  to  quantify the  large-scale  alignment  between the  matter
distribution in the Universe and the orientation of high density peaks
($M_{FoF}\geq10^{14}\hMsol$).   The  reference  samples are  generated
from 10\%  random subsets of  the overall particle  distributions.  If
real-space correlations  are considered  we use the  three dimensional
orientation   and    separation   vector   to    determine   $\theta$.
Alternatively,  if  observational consequences  are  discussed we  use
projected  orientations but still  3D distances  since the  effects of
reshift-space distortions are small.
\section{Results}
In this section we present our findings for the alignment
correlation functions in real- and redshift-space and for redshifts
z=0 and 0.6. 

Fig.~\ref{fig:z0} shows the conventional and alignment two point cross
correlation  functions  (upper  panels)  and  the residuals of the
alignment correlation functions about the conventional correlation
function (lower panels)  employing   different  lower  mass   cuts  ($1\times10^{14}$,
$5\times10^{14}$  and  $1\times10^{15}\hMsol$)  for the  high  density
peaks  in  the  main   sample.   Solid  lines  show  the  conventional
correlation   functions.  Dotted  and   dashed  lines   represent  the
correlation functions  along and perpendicular to  the orientations of
the   density  peaks.    The   gray  regions   represent  the   cosmic
variance. Interestingly,  the cosmic variance  in the lower  panels is
slightly smaller, this is because the differences between conventional
and alignment correlation functions  are independent of their absolute
values. To check  for systematic or numerical errors  we have repeated
the above analysis with randomly interchanged orientations within each
realization.  In this case we do not detect any significant difference
between the conventional and alignment correlation functions.

The  left  panel   of  Fig.~\ref{fig:z06}  displays  conventional  and
alignment  correlation  functions for density peaks  with masses larger
than   $5\times10^{14}\hMsol$  at   $z=0.6$.   The   right   panel  of
Fig.~\ref{fig:z06}  represents  the  correlation  functions  based  on
projected  orientations. Here,  the angles  used to  select  the pairs
along and perpendicular  to the orientations of the  density peaks are
computed between the projected orientations and the projected distance
vectors. Distances  are computed in  redshift space using  the distant
observer  approximation.    In  this  case  the   differences  in  the
amplitudes  between  the conventional  and  the alignment  correlation
functions are  somewhat reduced but  still significant beyond  the BAO
scale.

In conclusion, the alignment  correlation function between dark matter
density peaks with  masses  $\gtrsim10^{14}\hMsol$  and  the  overall  matter
distribution reveals anisotropies at scales larger than $150\hMpc$ far
beyond the BAO scale.  The difference between conventional correlation
functions and  correlation functions measured  along and perpendicular
to the orientations of the peaks is most pronounced for the most
massive peaks in real-space  at $z=0$.  But it remains  visible at
earlier  time and also if projected orientations are used.
\section{Discussion}
The findings presented here are based on N-Body simulations however  
qualitative similar effects are expected to shape the observed galaxy
correlation functions at scales up to $150\hMpc$. Potential
observational implications are:\\ 
  
$\bullet$ {\em  Large-scale anisotropy  and structure of two point
  correlation  functions:}   Our  results  indicate   that  matter  is
distributed anisotropically out to separations far larger then the BAO
scale.   At any  given  pair separation  the conventional  correlation
function is the average of the alignment correlation function over the
whole  range of  $\theta$.   We find  that  the alignment  correlation
function measured  along high density  peaks does not fall  below zero
(for separations $\le 150\hMpc$)  which is counterbalanced by negative
amplitudes at much  smaller separations for measurements perpendicular
to  it. At the BAO scale galaxy two point correlation
functions can be interpreted as the average of a more highly clustered
component along the direction of high density peaks and a less
clustered component perpendicular to it.\\
$\bullet$  {\em Shape and  location of the BAO peak:}  At BAO  scales the
amplitudes of the correlation functions between high density peaks and
the overall  matter distribution are significantly  higher if measured
along the orientation of the peaks.  In this case the BAO peak is composed
of a hump on top of the declining but still positive correlation
function.  The  signal perpendicular to the  orientations is dominated
by the BAO hump itself with negligible underlying clustering signal.  The
location of the BAO peak is found at somewhat smaller (larger) scales if
measured parallel  (perpendicular) to the orientation  of high density
peaks. For the measurements along
the density peak orientations the BAO hump is transformed into a plateau-like
feature.   If the  survey  volume is  dominated  by large  filamentary
structures the  shape of the conventional  correlation function should
be close to  that found for the parallel signal  shown here.  This may
relate to a study by \cite{Kazin-10} who find no apparent peak in 6\%
of their mock samples.\\
$\bullet$ {\em Zero crossing and large scale power:}
Several publications report the  non-detection of zero-crossing out to
scales of  $250\hMpc$ which  indicates (unexpected) large  scale power
\citep[e.g.,][]{Martinez-09,Kazin-10,Labini-09}.    We   find  similar
results for the correlation  functions measured along the orientations
of high  density peaks.  Furthermore, the  alignment cross correlation
function  measured along  the orientations  of high  density  peaks at
$z=0.6$ shows zero-crossing in contrast to the behavior at $z=0$.  The
WiggleZ    redshift-space     correlation    function    at    $z=0.6$
\citep{Blake-11}  shows  a  crossover  as  well.   Whether  these
analogies are coincidental remains to be explored in future
work.\\ 
$\bullet$ {\em Direct measurement of anisotropy:}
With the  advent of enormous  cluster catalogs \citep[e.g.,][]{Hao-10,
  Gilbank-11} it  should in principle be possible  to directly measure
the large  scale anisotropies with alignment  correlation functions if
cluster orientations  can be determined with  sufficient accuracy.  An
anisotropy signal  may even be extracted simply  by using orientations
of the cluster central  luminous red galaxies because the orientations
of    central    galaxies   and    host    systems   are    correlated
\citep{Faltenbacher-09, Okumura-Jing-Li-09, Schneider-11}.
\\
$\bullet$ {\em Improvement of BAO measurements:} 
If the  directional effects reported  here are observable it  would be
worthwhile considering  to measure the  BAO peak perpendicular  to the
orientations of  galaxy clusters (or  luminous red galaxies)  since in
this direction the BAO peak is better confined.
\section*{Acknowledgements} 
We would like to thank the anonymous referee for very helpful comments.
This  work  is sponsored  by NSFC  (no. 11173045), Shanghai Pujiang
Program (no. 11PJ1411600) and the CAS/SAFEA International Partnership
Program for  Creative  Research  Teams  (KJCX2-YW-T23). AF
acknowledges support from the South African SKA project. The
simulations used for this paper were performed on the Blade Centre
cluster of the Computing Center of the Max-Planck-Society in Garching,
and ICC Cosmology Machine COSMA4, which is part of the DiRAC Facility
jointly funded by STFC, the Large Facilities Capital Fund of BIS, and
Durham University. 
%
%

%
\end{document}